\documentstyle[twocolumn,prl,aps]{revtex}
\input{epsf.tex}
\def\inseps#1#2{\def\epsfsize##1##2{#2##1} \centerline{\epsfbox{#1}}}

\begin{document}
\def \beq{\begin{equation}}
\def \eeq{\end{equation}}
\def \beqarr{\begin{eqnarray}}
\def \eeqarr{\end{eqnarray}}

\twocolumn[\hsize\textwidth\columnwidth\hsize\csname @twocolumnfalse\endcsname

\draft

\title{
Spontaneous Breakdown of Translational Symmetry in Quantum Hall Systems:
Crystalline Order in High Landau Levels
}

\author{F. D. M. Haldane$^a$, E. H. Rezayi$^b$, and Kun Yang$^c$}

\address{
$^a$Department of Physics, Princeton University, Princeton, New Jersey 08544}

\address{
$^b$Department of Physics, California State University, Los Angeles, 
California 90032}

\address{
$^c$National High Magnetic Field Laboratory and Department of Physics,
Florida State University, Tallahassee, Florida 32306
}

\date{January 25, 2000, revised August, 2000}

\maketitle
\begin{abstract}
We report on results of systematic numerical studies of two-dimensional
electron gas systems
subject to a perpendicular magnetic field, with a high Landau
level partially filled by electrons. Our results are strongly suggestive of
a breakdown of translational symmetry and the 
presence of crystalline order in the
ground state. This is in sharp contrast with the physics of the lowest and
first excited Landau levels, and in good qualitative agreement
with earlier Hartree-Fock studies. Experimental implications of our results 
are discussed.
\end{abstract}

]

Recently there has been considerable
 interest in the behavior of a two-dimensional
(2D) electron gas subject to a perpendicular magnetic field, when a high
Landau level (LL) (with LL index $N\ge 2$) is partially filled 
by electrons. This is largely inspired by the recent experimental 
discovery\cite{lilly,du,cooper}
that the transport properties of the system are highly anisotropic and 
non-linear for LL filling fraction $\nu=9/2,11/2,13/2,\cdots$.
Previously, Hartree-Fock\cite{koulakov,fogler,moessner} (HF)
and variational studies\cite{fogler2} 
suggested that, unlike 
the $N=0$ and $N=1$ LL's (in which 
either incompressible fractional
quantum Hall (FQH) 
or compressible Fermi-liquid like states are realized), in $N\ge 2$ LL's the
electrons form charge density waves (CDW). In particular, at half-integral
filling CDW's break translational symmetry only in one direction and form 
stripes. Anisotropic transport would indeed result from such a striped (or 
related) structure\cite{fradkin,fertig,macdonald,phillips}.

We neglect  LL mixing, and consider the case where the LL with index  $N$
has partial filling $\tilde \nu$, while LL's with lower index are
completely filled ($\nu$ = $2N + \tilde \nu$). By particle-hole
symmetry of the partially-filled LL, this is equivalent to
$\nu$ = $2N + 2 - \tilde \nu$.   We also assume that the
partially-filled LL is maximally spin-polarized at the 
$\tilde \nu$ we consider.
Previously\cite{us}, we studied such  $N\ge 2$ LL's with
$\tilde \nu$ = 1/2 by
numerically diagonalizing the Hamiltonians of finite-size systems; 
those results strongly supported the existence
of stripe order. 

An outstanding issue is the nature of the ground state at high LL's
for fillings sufficiently far from the half-filled level.  Koulakov, Fogler and 
Shklovskii\cite{koulakov,fogler} (see also Moessner and Chalker\cite{moessner})
 predicted a novel crystalline phase
called the ``bubble'' phase with more than one electron per unit cell outside
of the range $\tilde\nu=0.4-0.6$.  The 
bubble crystal has lower energy than the Laughlin state for
$\nu=4+1/3$\cite{fogler2}.  Experimentally, a re-entrant quantum Hall state
is found near $\nu=4+1/4$ which is quantized as a $\nu=4$ LL 
plateau\cite{lilly,du,cooper}. Evidently the electrons in the 
top-most LL are frozen out of the transport. Pinning of a crystalline structure
provides a natural explanation of the re-entrant phase and would further 
explain the observed threshold in conduction\cite{cooper}.
However this is not entirely
conclusive and other mechanisms for the conduction threshold 
are also possible\cite{cooper}.

\begin{figure}
\inseps{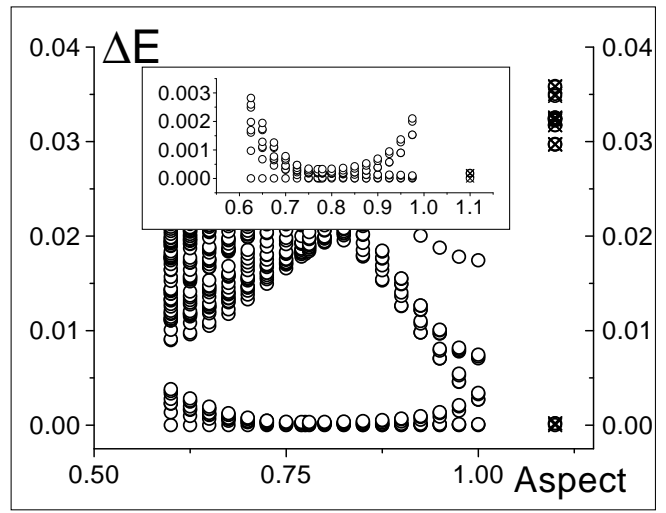}{0.36}
\caption{Energy levels versus aspect ratio for quarter-filled $N=2$ Landau level
with eight electrons and rectangular geometry. The inset is a blow-up 
of the low-energy spectra for aspect ratio between 0.6 and 1.0. 
The points at 1.1 (open circle plus x) correspond to hexagonal unit cell.}
\label{spectra}
\end{figure}

In this paper, we report on new numerical results on systems {\em away}
from half-filling using the unscreened Coulomb interactions. Remarkably, 
our results suggest that CDW's are formed at all filling factors we have 
studied, including those that would support prominent FQH
states or composite fermion Fermi-liquid states 
in the lowest or first excited Landau levels. These CDW's, however, 
have 2D structures and are no longer stripes when the filling factors are 
sufficiently far 
away from $1/2$. They are not Wigner crystals\cite{anderson} either,
unless $\tilde\nu$ is small (below 0.2).
In the intermediate filling factor
range, we find each unit cell of the CDW contains
more than one electron. Our results are in good agreement with the predicted
bubble phase and are the first exact finite-size calculations 
which exhibit
a crystalline state in a system with
continuous translational symmetry.

\begin{figure}
\inseps{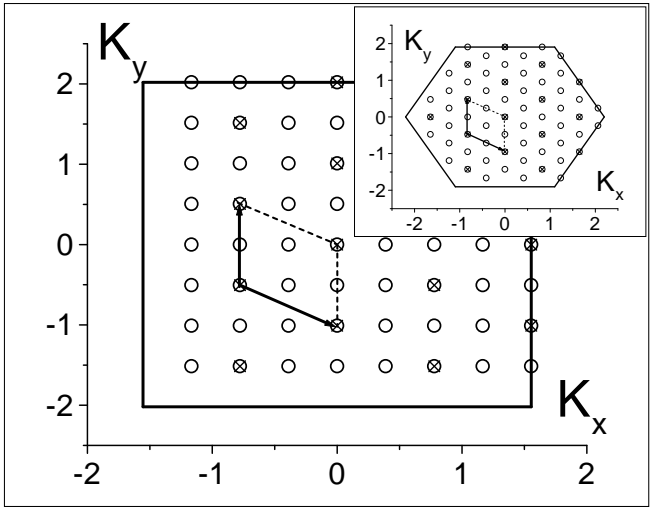}{0.36}
\caption{The allowed (circles) and the ground state manifold momenta
($\times$'s). The data is for a
rectangular geometry with $asp= 0.77$, 8 electrons 
in the $N=2$ LL and $\tilde\nu=1/4$. The solid line is the boundary of the
BZ. The superlattice reciprocal basis vectors are shown by solid arrows. The inset gives the corresponding results for an hexagonal unit cell.}
\label{momenta}
\end{figure}

We restrict the states of the electrons to a given LL, 
and work with periodic 
boundary conditions (PBC, torus geometry) as in our previous paper\cite{us}.
We also set the magnetic length to unity.
To detect intrinsically preferred configurations we consider a 
rectangular PBC unit cell and vary its aspect ratio.
The PBC plays a crucial role in removing continuous rotational symmetry,
and selecting a discrete set of possible crystal orientations.

In Fig.\ref{spectra}
we plot the energy levels of systems with $N_e=8$ electrons in the
$N=2$ LL at filling factor $\tilde\nu=1/4$ as a function of the aspect ratio.
We also show the levels of a system 
with a hexagonal PBC unit cell at the right side
of Fig.\ref{spectra}.
A generic feature of the spectra is the existence of
a large
number of low-lying states whose energies are almost degenerate, which we call
the ground state manifold. The momenta
of these quasi-degenerate states for rectangular geometry with aspect ratio
$asp=0.77$ and hexagonal geometry are shown in Fig.\ref{momenta}; 
they form a 2D superlattice
structure, which for the rectangular geometry  have the
super cell vectors ${\bf b}_1=2a\hat{e}_x-b\hat{e}_y$, and 
${\bf b}_2=2b\hat{e}_y$. Where,
$a=2\pi/L_1$, and $b=2\pi/L_2$. $L_1$, and $L_2$ are the dimensions 
of the unit cell ($L_1\times L_2=2\pi N_\Phi$, $N_\Phi$ is the total flux quanta
in the system).  The area per wavevector in the Brillouin zone (BZ) is $ab=(2\pi)^2/A$, where $A$
is the (real space) area of the system.

There are similarities as well as important differences between these spectra
and those of half-filled high LL's\cite{us} with stripe order. 
As in the stripe case\cite{us},
the large quasi-degeneracy of the ground state manifold is an 
indication of broken
translational symmetry\cite{note}. The difference here is
that (i) the degeneracy is much larger and (ii) the momenta of the low-lying
states form 2D instead of 1D arrays.
\noindent
These new features indicate that the translational 
symmetry is broken in both
directions and the ground 
state is a 2D CDW.
In the stripe state, on the other hand, the
translational symmetry is only broken in the direction perpendicular to the
stripes. Therefore the degeneracy is smaller and the momenta of the low-lying
states form a 1D array.

\begin{figure}
\inseps{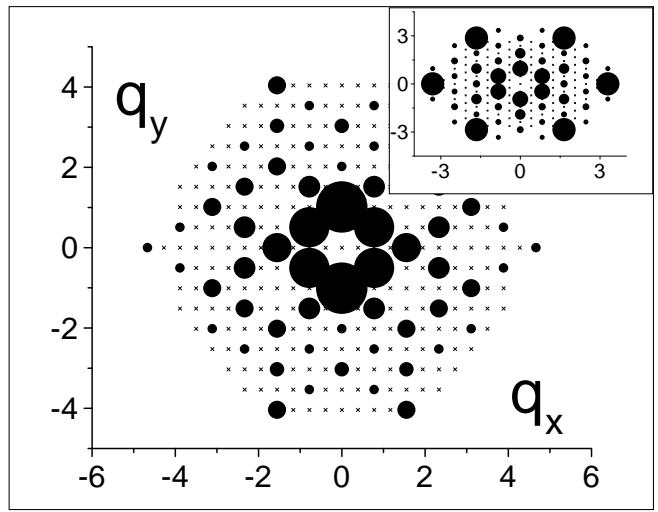}{0.36}
\caption{A 2D plot of the peaks of the
projected (or guiding center) charge susceptibility
$\chi({\bf q})$ at reciprocal lattice vectors, 
for a system with $N_e=8$ at $\tilde\nu=1/4$, rectangular 
geometry with $a=0.77$. The
size of the circles give an indication of the height of the peak
at that point. Only responses above 100 have been plotted as solid circles.
The zone boundaries are not within the range of the figure. 
The largest circle corresponds to 16491.
The inset gives the
results for a hexagonal unit cell. The $\times$'s are the allowed wavevectors.}
\label{chi}
\end{figure}
The momenta of the states in the ground state manifold are the
reciprocal lattice vectors of the bubble crystal. 
Transforming to the direct lattice vectors, we obtain ${\bf a}_1=\pi/a\hat{e}_x$ 
and ${\bf a}_2=\pi/2a\hat{e}_x+\pi/b\hat{e}_y$.  For the optimum system, with
$asp$=0.77, we obtain $a_1=8.08$, $a_2=7.42$, and $\phi=57^\circ$.  This is very
close to a triangular lattice.  In the case of hexagonal PBC unit cell, both
the reciprocal superlattice and its direct lattice are 
triangular.  

The number $N_D$ of distinct quasi-degenerate ground states allows the
number $N_b$ of bubbles in the system, and hence the number
$M$ = $N_e/N_b$ of electrons per bubble, to be immediately
obtained through the relation $N_bN_D$ = $\bar{N}^2$, where
$\bar{N}$ is the highest common  divisor  of $N_e$ and $N_{\Phi}$.
In our case, $\bar{N}$ = $N_e$ =  8, and  $N_D=16$, which gives 
$N_b=4$ and $M=2$\cite{foot1}.  The 
Wigner crystal would correspond to $N_b=N_e$ and $M=1$. 
In general\cite{duncan}, there are $\bar{N}^2$
distinct values of the total momentum quantum number,
which  define a BZ of area
$(2\pi\bar{N})^2/A$.  
If translational symmetry is 
broken, the area of the BZ of the superlattice is then
$(2\pi\bar{N})^2/AN_D$, which must be $(2\pi)^2/(A/N_b)$, where
$A/N_b$ is the area per bubble; hence
$N_bN_D$ = $\bar{N}^2$.

\begin{figure}
\inseps{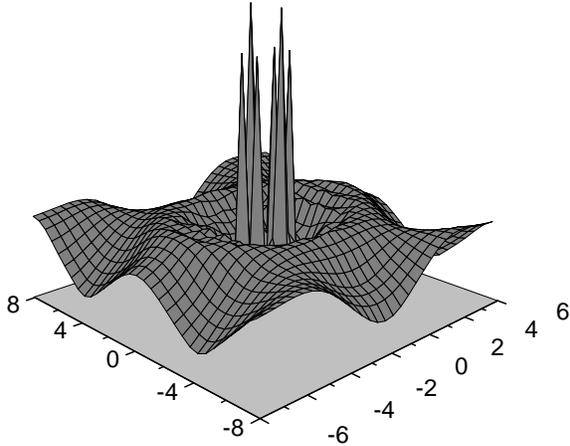}{0.36}
\caption{A 3D plot of the guiding center structure factor
$S_0({\bf q})$ (same system as Fig.\ref{chi}). The signature of the
hexagonal lattice is seen in the near six-fold symmetry of the peaks.}
\label{sq}
\end{figure}

We next turn to the density response functions.
In Fig.\ref{chi} we show the projected ground state charge 
susceptibility\cite{us} $\chi({\bf q})$ of one of the optimum rectangular and
hexagonal systems described above.
The calculation takes into account the contributions from the two lowest energy
states in each symmetry subspace; this is an excellent approximation in view of
the fact that 
the response function is dominated by low
energy states because of the energy denominator. 
We note that $\chi({\bf q})$ exhibits a strong response at the 
reciprocal lattice vectors (Bragg condition); 
the background at other wave vectors 
(shown by $\times$'s in Fig.\ref{chi}) are negligible
compared to these responses. The origin of the strong response lies in the 
approximate degeneracy among the states forming the ground state manifold. 
The system responds very strongly to a potential modulation with a wave vector 
that connects the ground state to one of the low lying  states (which must 
be a reciprocal lattice vector) because of the small energy denominator.
This is also another reason why there must be one low-lying state for each 
reciprocal lattice vector.
A second notable feature is the 
almost {\em hexagonal}
symmetry of the response, despite the fact
that the PBC geometry used in this case was {\em rectangular}. 
This indicates that the bubbles tend to form a triangular lattice,
in agreement with the predictions of HF theory.

The tendency toward forming a triangular lattice is also seen in the 
``guiding center (GC) static structure factor''
$S_0({\bf q})$\cite{us}, 
which we present as a 3D plot in Fig.\ref{sq}. Here we see 
sharp peaks with an approximate six-fold 
symmetry 
at the primary reciprocal lattice vectors,
indicating the presence of strong density
correlation at these wave vectors in the ground state.

\begin{figure}
\inseps{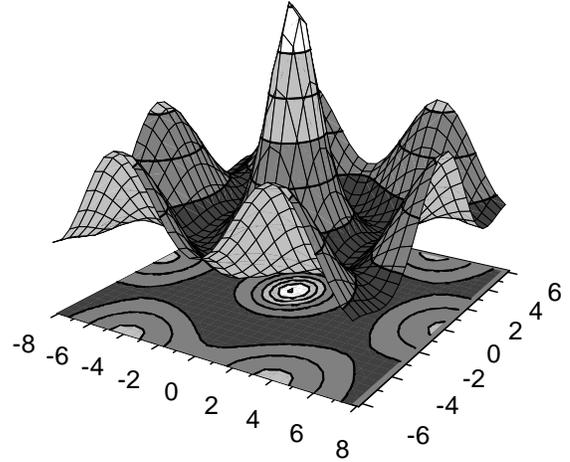}{0.36}
\caption{Real-space 
``projected density'' (guiding center) correlation function, 
derived from Fig.\ref{sq}.
A 2D contour plot is also included below the 3D plot.}
\label{guidepair}
\end{figure}
\begin{figure}
\inseps{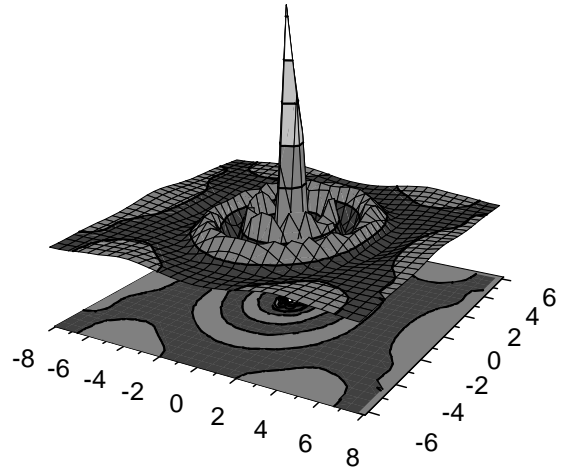}{0.36}
\caption{$N = 2$ Landau level full electron density correlations
(relative to a guiding center), derived from Fig.\ref{sq}.}
\label{elecpair}
\end{figure}

In Fig.\ref{guidepair}
and Fig.\ref{elecpair} 
we plot ground state ``projected density'' correlation functions 
in {\it real} space. 
These describe correlations relative to the GC 
({\it not} the coordinate) of a particle.
The first is the Fourier
transform (FT) of $S_0({\bf q})\exp (-q^2/ 2 )$, which is 
the electron density of an equivalent lowest-LL system, and gives
information on the spatial distribution of GC's.
The second is 
the FT of $S_0({\bf q})[L_N(q^2 /2)]^2 \exp (-q^2/ 2)$
with $N=2$ ($L_N$ is a Laguerre polynomial): this
(plus the uniform density of the filled LL's) represents the
actual electron density.

In Fig.\ref{guidepair}
the presence of four bubbles and the 
relative orientation of the bubbles can be clearly seen and 
there is strong crystalline order of the GC distribution.
The central peak contains two electrons, one of which is the particle
with the GC at the origin.
For $N > 0$, as in Fig.\ref{elecpair}, 
only weak order is displayed by the actual electron density, because
of the averaging effect of the cyclotron motion around the GC's.
It is the {\em guiding 
centers} of the electrons that form bubbles as 
anticipated in reference 7 (Fig.\ref{spectra}).
The electrons themselves manage to stay apart to
lower the Coulomb repulsion, in spite of the clustering of their
GC's.

\begin{figure}
\inseps{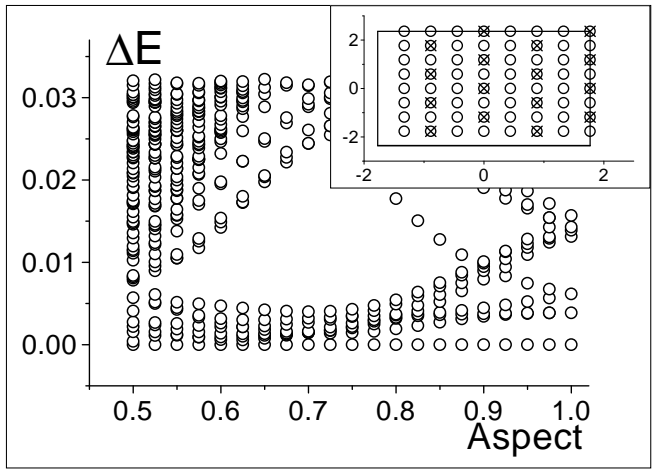}{0.36}
\caption{Spectra of systems with eight electrons at $\tilde\nu=1/3$ in the $N=2$ LL,
with rectangular geometry and various aspect ratios. The inset plots the
momenta of the low-lying states for $asp$ = 0.75, in the same way as Fig.
\ref{momenta}.}
\label{onethird}
\end{figure}

We have also explored other filling factors in the $N=2$ LL such as
$\tilde\nu=2/5$ and $\tilde\nu=1/3$, where
the system would
condense into prominent FQH states if it was in the lowest LL; 
here, however, our studies
suggest formation of CDW's instead. For 
$\tilde\nu=1/3$ we obtain  similar behavior to $\tilde\nu=1/4$:
the energy spectra 
as a function of the aspect ratio shown in Fig.\ref{onethird}
is very similar to 
Fig.\ref{spectra} and indicates formation of a 2D
CDW. Using the degeneracy of the 
ground state manifold, we find the number of electrons per bubble  
is also two.
The energies of the states in the ground state manifold, however, are not as
close as the $\tilde\nu=1/4$. This results in weaker peaks in $\chi({\bf q})$ at
the reciprocal lattice vectors. 

We interpret this to be an indication that,
in this
LL, $\tilde\nu=1/4$ is more favorable than $\tilde\nu=1/3$
for formation of a two-electron
bubble phase. 
In real systems a crystal is always pinned by a
disorder potential, and in a nonlinear transport measurement, there should be
a threshold depinning 
field at which there is a sharp feature in the $I-V$ curve. 
A weaker crystal would result in more diffuse 
conduction threshold as various portions of the crystal get depinned at 
different current values, while a stronger one, on the other hand, will
have  sharp conduction threshold.  
This is consistent with the observation of 
Cooper {\sl et al}\cite{cooper} that there is a
sharp threshold region 
at about $\tilde\nu$ = 1/4, but more diffuse thresholds at
both higher and lower $\tilde \nu$.

In contrast to $\tilde\nu=1/4$ and $\tilde\nu=1/3$, 
the spectra for $\tilde\nu=2/5$ was found to be
very similar to $\tilde\nu=1/2$\cite{us}.  The momenta of the low-lying 
states belong to a 1D array, indicating formation of a 1D CDW or stripe phase;
the weight of the HF state 11110000001111000000 ($N_e=8$), in  a rectangular 
geometry with an aspect ratio of 0.80, is about 65\%. 
We  conclude that the transition from stripe to bubble 
phases occurs between $\tilde\nu=1/3$ and $\tilde\nu=2/5$, in qualitative agreement
with HF predictions\cite{fogler,moessner}. We have also studied higher LL's.
The results are similar and will be reported elsewhere.  

We have benefited from stimulating discussions with J. P. Eisenstein,
K. B. Cooper and M. P. Lilly.  We thank B. I. Shklovskii and M. M. Fogler
for helpful comments.
This work
was supported by NSF DMR-9420560 and DMR-0086191 (E.H.R.), DMR-9809483 (F.D.M.H.),
DMR-9971541, and the Sloan Foundation (K.Y.). E.H.R. acknowledges the
hospitality of ITP Santa Barbara, supported by NSF-PHY94-07194,
 where part of the work was performed.


\begin{references}
\bibitem{foot1} This result is in overall agreement with the estimate 
of reference 5: $M=3\tilde\nu N$.
\end{references}
\end{document}